\documentclass{article}
\usepackage{amsfonts}
\usepackage{graphicx} 
\usepackage{amsmath}
\usepackage{natbib}
\usepackage{booktabs}
\usepackage{subcaption}
\usepackage{wrapfig}  
\usepackage{multirow}
\usepackage{authblk}
\usepackage{hyperref}
\DeclareMathOperator*{\argmin}{argmin}
\newcommand{\sign}{\operatorname{sign}}

\title{Univariate-Guided Sparse Regression for Biobank-Scale High-Dimensional Omics Data}

\author[1]{Joshua Richland}
\author[2]{Tuomo Kiiskinen}
\author[3]{William Wang}
\author[1]{Sophia Lu}
\author[1,2]{Balasubramanian Narasimhan}
\author[1,2]{Trevor Hastie}
\author[2]{Manuel Rivas}
\author[1,2]{Robert Tibshirani}

\affil[1]{Department of Statistics, Stanford University}
\affil[2]{Department of Biomedical Data Science, Stanford University}
\affil[3]{Department of Computer Science, Stanford University}

\begin{document}

\maketitle

\begin{abstract}
We present a scalable framework for computing polygenic risk scores (PRS) in high-dimensional genomic settings using the recently introduced Univariate-Guided Sparse Regression (uniLasso). UniLasso is a two-stage penalized regression procedure that leverages univariate coefficients and magnitudes to stabilize feature selection and enhance interpretability. Building on its theoretical and empirical advantages, we adapt uniLasso for application to the UK Biobank, a population-based repository comprising over one million genetic variants measured on hundreds of thousands of individuals from the United Kingdom. We further extend the framework to incorporate external summary statistics to increase predictive accuracy. Our results demonstrate that uniLasso attains predictive performance comparable to standard Lasso while selecting substantially fewer variants, yielding sparser and more interpretable models. Moreover, it exhibits superior performance in estimating PRS relative to its competitors, such as PRS-CS. Integrating external scores further improves prediction while maintaining sparsity.
\end{abstract}

\section{Introduction}
\label{sec: Introduction}

With rapid advancements in computing power and storage capacity, data collection has become increasingly detailed and large in scale. As a result, scientists routinely encounter high-dimensional data settings, where the number of predictors is much larger than the number of observations ($p \gg n$). Such scenarios are common in fields like genomics and finance, where granular information can be recorded for each individual or transaction. In these regimes, regression has become a central tool for both prediction and variable selection, underpinning a wide range of modern applications.
A second defining feature of -omics design matrices, in particular, is strong correlation among predictors. In genomics, linkage disequilibrium (LD) creates blocks of SNPs that carry very similar information; in transcriptomics, co-expression and shared regulatory programs induce clusters of correlated genes. A truly causal signal at one locus therefore tends to bring along a cloud of correlated predictors with similar univariate behaviour. Any multivariate method must decide what to do with these groups—either select a sparse set of representatives, or distribute (“smear”) the effect across many correlated variables.
Although our main focus is on -omics applications, similar large $n$, larger $p$ structures arise in other domains. In finance, for instance, one may observe long time series or many individual transactions ($n$) together with hundreds or thousands of asset-, sector-, and macro-level predictors ($p$), many of which move together through shared risk factors or market regimes. High-dimensional sparse regression in such settings faces the same core challenges: correlated predictors, the need to identify a small subset of informative signals, and stringent computational constraints.

Often, only a much smaller subset of features have a meaningful relationship with the response. Thus a main goal of researchers is to focus on identifying which variables are worth the investment of time and resources to measure and analyze. 
\subsection{Review of the Lasso}

The Lasso \citep{Tibs1996} performs both estimation and variable selection and has become a cornerstone of modern statistics and machine learning.

For a model matrix $X \in \mathbb{R}^{n \times p}$ and response vector $y \in \mathbb{R}^n$, Lasso solves the following minimization problem:

\begin{equation}
\hat{\beta} = \argmin_{\beta} \; \frac{1}{2n} \| y - X\beta \|_2^2 + \lambda \| \beta \|_1.
\end{equation}
$\lambda$ is a regularization parameter that determines the number of features identified in the solution set. In practice, we evaluate the objective over a grid of $\lambda$ values and then select the best $\lambda$ using either cross-validation or performance on a held-out validation set. Equivalently, we can formulate the Lasso problem in the following way:
\begin{equation}
\hat{\beta} =  \argmin_{\beta} \; \frac{1}{2n} \| y - X\beta \|_2^2 
\quad \text{subject to} \quad \| \beta \|_1 \leq t
\end{equation}
where $t>0$ is a tuning parameter controlling the sparsity of the solution, directly related to $\lambda$ in the Lagrangian formulation. The constraint sets an upper bound on how large the coefficients $\hat{\beta}$ can be in total (in terms of their $\ell_1$ norm, which is $\sum_{j=1}^{p} |\beta_j| $). A larger $t$ allows the coefficients to grow by imposing a smaller amount of regularization.

\subsection{Optimization Approaches for the Lasso}
A variety of computational tools have been developed to efficiently solve Lasso problems. \textbf{\tt glmnet} \citep{glmnet} is among the most widely cited R packages, capable of computing entire solution paths for a range of generalized linear models. \textbf{\tt snpnet} \citep{Qian2020} is another R package designed specifically for large-scale, high-dimensional Lasso regression in genomics applications. \textbf{\tt snpnet} stands out as one of the most competitive solvers, combining strong-rule screening with efficient KKT batching to scale to millions of SNPs. This approach yields high predictive accuracy in large-scale genomic analyses.

Elastic Net (EN) is another path solver that extends the framework by incorporating an $\ell_2$ penalty, thereby combining sparsity with ridge regularization. This generalization provides greater flexibility and stability in settings with correlated features, where pure $\ell_1$ regularization may perform suboptimally. EN seeks to minimize the following objective function:

\begin{equation}
\hat{\beta} = \argmin_{\beta} \left\{
\frac{1}{2n} \| y - X\beta \|_2^2
+\lambda \left(
\alpha |\beta|_1
+(1 - \alpha) |\beta|_2^2
\right)
\right\}.
\end{equation}
Here, $\lambda > 0$ controls the overall regularization strength, while $\alpha \in [0,1]$ balances the contribution of the $\ell_1$ penalty and the $\ell_2$ penalty. Setting $\alpha = 1$ recovers the Lasso, while $\alpha = 0$ yields ridge regression. This unified structure allows practitioners to interpolate between the two classical approaches within a single flexible framework.


%

 Adelie \citep{adelie2024} 
is a novel Python package (with a version in R as well) designed for efficiently solving lasso and group lasso problems. This package can handle large-scale inputs with greater computational efficiency than traditional solvers, and the package provides a suite of generalized linear model (GLM) classes. Benchmarks against other state-of-the-art solvers demonstrate that \textbf{\tt Adelie} achieves faster speeds than other group lasso packages with equivalent accuracy. On Lasso and EN problems, it performs comparably to the \textbf{\tt glmnet} R package. Additionally, \textbf{\tt Adelie} offers flexible parameter tuning, warm-start capabilities, and easy integration with standard Python data science workflows, making it powerful for high-dimensional regression tasks.

Another notable feature of \textbf{\tt Adelie}, similar to \textbf{\tt snpnet}, is its ability to efficiently fit lasso and EN models on genome-wide association study (GWAS) datasets in Python. To handle the scale and complexity of such data, \textbf{\tt Adelie} introduces specialized file formats optimized for both memory efficiency and the speed of matrix operations. This design allows the package to process large-scale genotype and phenotype matrices that would otherwise exceed standard memory limits, enabling scalable analyses of GWAS data.

\subsection{The UK Biobank Database}
The UK Biobank \citep{UKBiobank} provides an excellent testbed for scientists using such regression techniques to investigate quantities of interest, such as the variance explained by genomic predictors, and to identify SNPs potentially associated with specific traits. The UK Biobank is a large, population-based repository containing genetic and health information from approximately 500,000 individuals across the United Kingdom. For each participant, it provides data on over 1,000,000 single nucleotide polymorphisms (SNPs) alongside a wide range of phenotypes. 

The sheer scale of the UK Biobank introduces substantial computational challenges for analyzing and sparsifying the dataset. Efficient storage and manipulation of this high-dimensional data are pivotal for minimizing objective functions in regression and other statistical models. Recent advances in algorithms and computational tools have made large-scale analyses of datasets of this magnitude increasingly feasible.

Regression provides a natural framework for studying the UK Biobank data, allowing researchers to quantify relationships between SNPs and phenotypes. Packages such as \textbf{\tt Adelie} and \textbf{\tt snpnet} enable analyses to be performed efficiently on such large-scale datasets. By incorporating regularization techniques, i.e. Lasso, regression can produce sparse solutions, effectively reducing the set of over a million SNPs to a smaller subset most likely to be associated with the response. This not only aids in interpretability, but also improves predictive performance by mitigating overfitting in high-dimensional settings.

Polygenic risk scores (PRS) summarize genetic liability as a weighted sum of genome-wide variants. At biobank scale, the standard PRS workflow is two-steps: first, perform a univariate genome-wide association study (GWAS) to obtain marginal effect estimates for each SNP; second, apply an LD-aware shrinkage or selection method (for example, clumping and thresholding, LDpred, or PRS-CS) to convert these marginal effects into a genome-wide predictor. Historically, even this first step already pushed computational limits (fitting millions of univariate regressions on hundreds of thousands of individuals) so genome-wide multivariate regressions were not practically feasible at scale. This computational constraint, together with the complexity of modelling LD explicitly, is a major reason why univariate GWAS has remained the dominant analysis framework for so long. Because GWAS treats each SNP in isolation, a true signal within an LD block typically lifts an entire neighbourhood of correlated SNPs, and without strong shrinkage the naive sum of genome-wide GWAS betas would imply more variance explained than is empirically possible. The second step exists precisely to correct for this double-counting and sampling noise, but it usually leads to dense predictors with highly flattened per-SNP effect sizes.

\subsection{Contributions}
In this paper, we implement a novel method for analyzing ultra-high-dimensional -omics data that directly fits a multivariate model on individual-level data, rather than relying on the standard two-step GWAS-plus-PRS pipeline.  We present an adaptation of Univariate-Guided Sparse Regression (uniLasso) \citep{uniLasso} tailored for biobank-scale SNP data. Empirically, uniLasso achieves predictive performance comparable to existing methods but with ~40\% fewer predictors, significantly improving the interpretability of the results. 

Additionally, we consider the integration of external data sources to further enhance predictive accuracy. In biomedical research, privacy regulations often prevent the sharing of raw data, so researchers sometimes may only provide summary statistics rather than entire data sets. Following the approach proposed in \cite{uniLasso}, we demonstrate how to incorporate external scores from independent datasets into the uniLasso pipeline, bolstering polygenic risk score (PRS) predictions. We show how these external scores can be combined with individual-level data in a single sparse multivariate model, rather than as a separate post-GWAS step.

This paper is organized as follows: Section \ref{sec: proposal} introduces our adaptation of the uniLasso algorithm, beginning with a brief recap of the standard method before describing how we extend it to both continuous and binary response variables in ultra-high-dimensions. In Section \ref{sec: Biobank analysis} we then demonstrate its practical use through an application to the UK Biobank. We outline how to incorporate external scores and present additional experiments highlighting their benefits in Section \ref{sec: external scores}. Section \ref{sec: comparison summary} summarizes comparisons of uniLasso to other approaches for constructing polygenic risk scores. Finally we outline future challenges in Section \ref{sec: discussion}.

\section{Proposed Method}
\label{sec: proposal}
\subsection{Overview of the uniLasso Algorithm}

Univariate-Guided Sparse Regression (uniLasso) \citep{uniLasso} is a two-stage regression framework that promotes sparsity and interpretability by leveraging information from univariate regressions. Specifically, uniLasso constrains the coefficients in a multivariate model to align with the signs of their corresponding univariate coefficients, enhancing predictive accuracy and improving the interpretability of selected predictors. 

The uniLasso procedure, as outlined in \citet{uniLasso}, assumes the same supervised learning framework as Lasso. For a model matrix $X \in \mathbb{R}^{n \times p}$ and response vector $y \in \mathbb{R}^n$, the uniLasso procedure is defined as follows:

\begin{enumerate}


\item  For $j=1,2,\ldots,p$ compute the univariate intercepts and slopes $(\hat\beta_{0j}, \hat\beta_j)$ and the leave-one-out (LOO) fitted values
\[
\hat y^{-i}_{ij} \;=\; \hat\beta_{0j}^{-i} + \hat\beta_j^{-i} x_{ij}, 
\qquad i=1,\ldots,n,\; j=1,\ldots,p,
\]
and collect them in a new $n\times p$ feature matrix $F$ with entries $F_{ij} = \hat y^{-i}_{ij}$.

\item  Fit a non-negative lasso with an intercept and no standardization to target $y$ using the LOO univariate predictions in $F$ as features:
\begin{align*}
&\argmin_{\theta} \quad 
 \left\{
  \frac{1}{n} \sum_{i=1}^n
  \Bigl(
    y_i - \theta_0 
    - \sum_{j=1}^p 
      F_{ij}\, \theta_j
  \Bigr)^2
  + \lambda \sum_{j=1}^p \theta_j
  \right\} \\
&\text{subject to} \quad 
 \theta_j \ge 0,\quad j=1,\ldots,p.
 \ \text{Select $\lambda$ through cross validation.}
\end{align*}

\item The final model can be written as  
$$ \hat\eta(x)=\hat\gamma_0 + \sum_{j=1}^p\hat\gamma_j x_j,$$
with $\hat\gamma_j=\hat\beta_j\hat\theta_j$, and $\hat\gamma_0=\hat\theta_0+\sum_{\ell=1}^p\hat\beta_{0\ell}\hat\theta_\ell$.
\end{enumerate}
Note that we do not standardize the features before applying the non-negative Lasso in Step 2 because the entries of $F$ are all on the scale of the response. Closed-form expressions for the LOO fits and the univariate coefficients are provided in Section \ref{sec: adapted algorithm}.

The steps taken in this procedure serve several purposes. First, it ensures that the signs of the final coefficients agree with the signs of the univariate coefficients (or they are zero). This increases the interpretability of the result, as it excludes features whose coefficient sign changes between the univariate and multivariate model. In addition, features with larger univariate coefficients will also tend to have larger coefficients in the final model, since they are given an advantage in the penalization of $\theta$. Moreover, using LOO fitted values rather than raw univariate estimates or the original features amounts to feeding cross-validated univariate predictions into the Lasso in Step~2. This forces the selection step to rely on out-of-sample predictive signal rather than in-sample noise, which empirically leads to sparser models and improved generalization \citep{uniLasso}. It also justifies the use of cross validation.

As demonstrated in the uniLasso paper, this method has exhibited excellent performance, producing sparser solutions than classical techniques. In its original formulation, however, uniLasso is not immediately scalable to biobank-sized datasets, which prevents a direct comparison with genomics tools such as \textbf{\tt snpnet}. 

One challenge posed by uniLasso in ultra-high-dimensional data analysis is the need to store the immense feature matrix. In computation, Step~1 of the uniLasso pipeline actually requires the user to remember another $(n \times p)$ matrix of the LOO fitted values. While SNP matrices are typically sparse (many of the entries are 0), the LOO matrix $F$ is not. Thus the computer must be capable of storing another feature matrix which could take up over 1500 GB of memory.

Computation time arises as a second challenge when calculating the univariate coefficients and LOO fitted values. Since they need to be computed for each column, when the number of covariates is in the order of millions (as is the case in -omics settings) this procedure can take a long time.

A final challenge arises in Step~2 of the algorithm when choosing the regularization value. Again, given the scale of the data, cross validation becomes computationally infeasible. This is because each of the $k$ folds requires refitting the model on a large proportion of the entire dataset, leading to prohibitively large computational, memory, and time costs.

\subsection{Details of the Omics-adapted uniLasso Algorithm} \label{sec: adapted algorithm}

To address the computational challenges discussed above, we propose a modified version of the uniLasso algorithm designed for application in general -omics settings. We demonstrate that this method achieves predictive performance comparable to standard tools such as \textbf{\tt snpnet}, while producing substantially sparser and more interpretable solutions.

We use Sherlock, a high-performance computing (HPC) cluster at Stanford University, to engineer the uniLasso pipeline using the \textbf{\tt Adelie} Python package. Our compute instances use a cluster of Intel(R) Xeon(R) E5-2640 v4 CPUs with 64 total cores. This setup enables efficient allocation of memory and computational resources necessary to sustain the full analysis. When calculating the leave-one-out (LOO) fits and univariate coefficients, it is important to note that each computation depends only on individual columns of the design matrix. 

For a given SNP column $x$, let $\tilde{x} = x - \bar{x}$, where $\bar{x}$ denotes the column mean. Similarly, define $y$ as the response and  $\tilde{y} = y - \bar{y}$ as its centered counterpart. The univariate slope $u_1$ and intercept $u_0$ are then given by:
\[
u_1 = \frac{\tilde{x}^\top \tilde{y}}{\| \tilde{x} \|^2}, 
\hspace{2cm} 
u_0 = \bar{y} - u_1 \cdot \bar{x}.
\]
The vector of LOO fitted values $\vec{f}$ corresponding to the column $x$ are defined as follows, where $H_{ii}$ denotes the $i^{th}$ diagonal element of the hat matrix $H$:

\[
y_i - f_i = \frac{y_i - (u_0 + u_1 x_i)}{1 - H_{ii}}, 
\quad 
\text{where } 
H_{ii} = \frac{1}{n} + \frac{\tilde{x_i}^2}{\| \tilde{x} \|^2}.
\]
Using vectorized formulas suggested in \citet{uniLasso} we parallelized the computations, which occur in an embarrassingly parallel manner. This allowed us to fully exploit CPU resources and parallel computing to markedly accelerate computation.

In the standard uniLasso algorithm, the regularization parameter in Step~2 is typically chosen via $k$-fold cross validation. However, once again given the scale of our data, this is computationally infeasible. Instead, following the approach of \citet{Qian2020}, we employed a train-validation-test split (70/10/20) to efficiently tune the regularization strength.
Specifically, we first fit the model on the training set over a grid of $\lambda$ values. Using the validation set, we then identified the value of $\lambda$ that achieved the best predictive performance. Next, we refit the model on the combined training and validation sets with the chosen parameter. Finally, we evaluated the resulting model on the held-out test set to obtain the optimal coefficients. 

This validation-based approach substantially reduced computational time and the need to store large amounts of data, while yielding empirically strong results. We also believe that this strategy is an effective and generalizable approach for hyperparameter tuning in ultra–high-dimensional settings.

Moreover, SNP data is known to be particularly sparse, typically taking values in the set $\{0, 1, 2\}$. Typically, a majority of the entries in a given column of a SNP feature are 0. To address the small levels of variance within certain column, we implemented two strategies:

\begin{enumerate}
    \item \textbf{Minor allele frequency (MAF) filtering:} \\
    Minor allele frequency (MAF) is the proportion of the less common allele among all observed alleles at each SNP. SNPs with MAF below 0.0005 were excluded prior to applying uniLasso (or Lasso). This filtering step is standard in genomic analyses, as rare variants tend to contribute little predictive power and can lead to unstable coefficient estimates. 
    
    \item \textbf{Stabilizing regularization of low-variance columns:} \\
    For SNPs with MAF $\geq 0.0005$, small variances still occasionally caused univariate coefficients to explode, leading to unstable estimates. One key quantity that appears in the denominator of the univariate and LOO calculations is 
    \[
        S_{xx} = \tilde{x}^\top \tilde{x}, \quad \text{where } \tilde{x} = x - \bar{x}.
    \] 
    \citet{Tusher2001} encountered a similar instability in the context of microarray analysis, where small denominators led to highly variable test statistics. They propose adding a data-dependent constant to stabilize the denominator. Following this approach, we regularize low-variance SNP columns by replacing $S_{xx}$ with 
    \[
        S_{xx}^* = S_{xx} + \text{fifth-percentile}(S_{xx}),
    \]
    where the additive constant corresponds to the fifth percentile of the empirical distribution of $S_{xx}$ values over all columns of the SNP matrix. This adjustment prevents unstable estimates arising from sparse or near-constant genotype columns when $S_{xx}$ approaches zero.
\end{enumerate}

Both modifications substantially improved stability across data partitions and, in most cases, even enhanced predictive performance relative to non-regularized runs.

\subsection{Omics-adapted uniLasso Algorithm for Binary Response Variables}

In the case where the response variable is binary rather than continuous ($y_i \in \{0,1\}$ for all $i$), we use a penalized logistic regression objective function:

\begin{align*}
\hat{\beta} 
= &\argmin_{\beta} 
\left\{
-\frac{1}{n}\sum_{i=1}^{n} 
\left[
y_i \log p_i + (1 - y_i)\log(1 - p_i)
\right]
+ \lambda \sum_{j=1}^{p} |\beta_j|
\right\}, \\
&\text{where } 
p_i = \frac{\exp(\beta_0 + x_i^\top \beta)}{1 + \exp(\beta_0 + x_i^\top \beta)}.
\end{align*}
Similar to the Lasso, we introduce regularization to control the sparsity of the solution.

For uniLasso, recall that the algorithm requires LOO fitted values and univariate coefficients. Unlike the case with Gaussian loss, there are no closed-form solutions for these expressions. However, in practice, logistic regression is typically fit via iteratively reweighted least squares (IRLS). We follow the procedure described in \citet{uniLasso}, summarized below:

Given the fitted linear predictor vector $\eta_j^{(\ell)}$ at iteration $\ell$, one forms a working response vector $z_j^{(\ell)}$ that depends on $\eta_j^{(\ell)}$, $y$, and the GLM family, along with an observation weight vector $w_j^{(\ell)}$. The updated model $\eta_j^{(\ell+1)}$ is then obtained by performing weighted least squares of $z_j^{(\ell)}$ on $x_j$ with weights $w_j^{(\ell)}$. In our implementation, we perform two IRLS iterations per feature.

As in the Gaussian setting, we parallelize and vectorize operations to efficiently fit all $p$ univariate logistic regression models. The expressions are only slightly more complex than in the unweighted case. For the LOO fits, we adopt the approximation proposed by \citet{RadMaleki2020}, which provides estimated LOO predictions without refitting and has been shown to work well in large-scale GLM applications. This corresponds to using the final weighted least squares IRLS iteration from each univariate fit when fitting the univariate models. We use the formula they propose which accommodates observation weights. 

Uniquely, we apply the same regularization and MAF filtering when running the binary uniLasso algorithm. Aside from the modifications required for the logistic loss and the corresponding univariate values, the majority of the uniLasso pipeline remains nearly identical to the Gaussian case.

\section{UK Biobank Analysis}
\label{sec: Biobank analysis}
We describe a real-data application on the UK Biobank that demonstrates the utility of the uniLasso algorithm.

The UK Biobank \citep{UKBiobank} is a large prospective cohort study containing extensive genetic and phenotypic data from approximately 500,000 middle-aged individuals across the United Kingdom. In this analysis, we focus on modeling the relationship between individual genotype sequences and their corresponding phenotypic traits. Using uniLasso, we aim to identify SNPs that are most informative for predicting the phenotype of interest. Our proposed adaptation enables model fitting at full scale and yields a reduced active variable set compared to alternative methods. This smaller collection of selected SNPs enhances interpretability while maintaining comparable levels of variance explained.

We restricted our analysis to a subset of 336,442 White British individuals from the full UK Biobank dataset who satisfy the population stratification criteria described in \citet{subset_UKBiobank}. The data were randomly partitioned into training ($70\%$), validation ($10\%$), and test ($20\%$) subsets. Each individual has 1,080,968 measured genetic variants, where each variant is encoded by one of three levels: 0 for homozygous major alleles, 1 for heterozygous alleles, and 2 for homozygous minor alleles. Missing genotypes were imputed using the mean of the observed values. In addition to genotype data, covariates such as age, sex, and 10 precomputed principal components of the SNP matrix were included to account for population structure. Among the thousands of measured phenotypes in this dataset, the phenotypes we used to evaluate our method were standing height, body mass index (BMI), coronary heart disease (CHD), and asthma. CHD and asthma are binary responses, where 1 indicates disease and 0 no disease.

\begin{table}[t]
\centering
\begin{tabular}{lcccc}
\toprule
\multicolumn{5}{c}{\textbf{Number of Non-Zero Coefficients}} \\
\midrule
\textbf{Model} & \textbf{Height} & \textbf{BMI} & \textbf{CHD} & \textbf{Asthma} \\
\midrule
Lasso       & 55,038 & 33,076 & 2,615 & 5,556 \\
uniLasso    & 34,256 & 18,833 & 1,009 & 3,030 \\
uniLasso ES & 44,788 & 24,598 & 1,310 & 5,698 \\
\bottomrule
\end{tabular}
\caption{\em Number of non-zero coefficients for each phenotype measured across all Lasso-based algorithms. Compared to Lasso, uniLasso selects 38\% fewer SNPs for height and 43\% fewer for BMI, while only reducing $R^2$ by 0.006 and 0.016, respectively. uniLasso with external scores (uniLasso ES) is introduces in Section \ref{sec: external scores}.}
\label{tab:nonzero_coefs}
\end{table}

To evaluate predictive performance for the continuous responses, we use the $R^2$ statistic: given a linear estimator $\hat{\beta}$ and data ($y, X$), $R^2$ is defined as

$$
R^2 = 1 - \frac{\|y - X\hat{\beta}\|_2^2}{\|y - \bar{y}_0 \textbf{1}\|_2^2},
$$
where $\bar{y}_0$ is the average response of the training data and \textbf{1} is the $n$-vector of all ones. The PRS for individual $i$ is the linear predictor $\widehat{\mathrm{PRS}}_i = x_i^\top \hat\beta$, where $x_i$ denotes that individual’s vector of SNP genotypes. In this formulation, covariates enter only through the evaluation model used to predict the phenotype itself. The resulting $R^2$ quantifies the proportion of variance explained by this combined model. Ranging from 0 to 1, higher values represent better explanatory power by the model. 

For binary responses, we assess predictive performance using the receiver operating characteristic (ROC) curve, which illustrates the trade-off between true and false positive rates across different classification thresholds. The area under the ROC curve (AUC) reflects how well the model distinguishes between positive and negative outcomes, with values ranging from 0.5 (no discriminatory ability) to 1 (perfect discrimination). For phenotypes with binary outcomes, we evaluate the AUC for all regression methods used.

\begin{figure}[t] 
  \scalebox{0.6}{\includegraphics{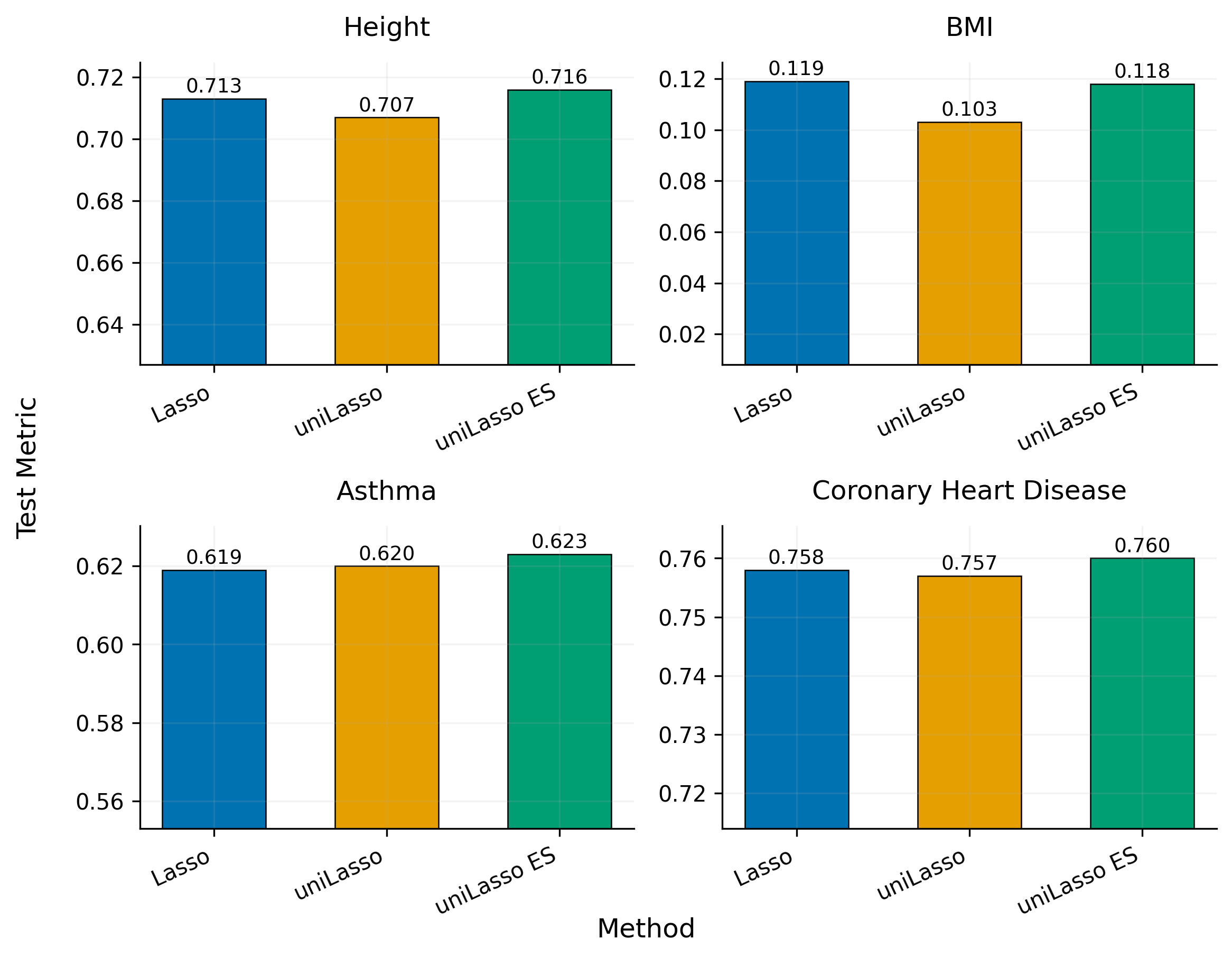}}
  \caption{\em{ Comparison of the test set predictive performance of the different polygenic risk score (PRS) methods with refitting on the training and the validation set.} All regressions were performed with Adelie in Python. Test-set $R^2$ is the metric evaluated for the continuous phenotypes (height and BMI), and AUC is evaluated for the binary phenotypes (asthma and CHD).}
  \label{fig:test errors}
\end{figure}

We perform Lasso and uniLasso on all of the aforementioned phenotypes to have a sense of any potential advantages. For the continuous phenotypes (height and BMI) there is a consistent trend: Lasso outperforms uniLasso by a small margin but uses a significantly larger number of coefficients in its solution. The small loss in $R^2$ is worthwhile for uniLasso when considering the nearly $40\%$ reduction in active-set size. Figure \ref{fig:test errors} illustrates the difference in their performance on held out test sets while Table \ref{tab:nonzero_coefs} displays the number of non-zero coefficients at the optimal $\lambda$ value for each method tested on each phenotype.

In the binary case, uniLasso and Lasso perform much more similarly as measured by AUC, with uniLasso slightly outperforming Lasso on Asthma. Otherwise, the difference in test error between Lasso and uniLasso does not exceed 0.016, an encouragingly small margin. Excitingly, as also seen with the continuous responses, uniLasso maintains a significantly sparser solution set than Lasso (See Table \ref{tab:nonzero_coefs}). For CHD, uniLasso is over 50\% sparser than Lasso and has lower test error by only 0.001. This reduced solution set in conjunction with the alignment of signs between univariate and multivariate coefficients reinforces the merits of uniLasso. Section \ref{sec: external scores} goes into more detail about an alternative formulation of uniLasso that potentially achieves a more optimal tradeoff.

Table \ref{tab: sign_changes} summarizes the number of sign changes observed between the univariate coefficients and the non-zero coefficients in final Lasso model. This metric is a reflection of the interpretability of the multivariate (Lasso) model: sign changes indicate that the multivariate fit has flipped the direction of association for certain variables. Such reversals are difficult to interpret because the coefficient’s sign becomes dependent on the presence of other variables. As shown in the table, the binary models exhibit far fewer sign changes than the continuous ones, which is expected given their smaller active sets. A built-in advantage of uniLasso is the lack of sign changes regardless of the type of response variable.

\begin{table}[t]
\centering
\begin{tabular}{lcc}
\toprule
\multicolumn{3}{c}{\textbf{Number and Percentage of Sign Changes in Lasso}} \\
\midrule
\textbf{Phenotype} & \textbf{\# Sign Changes} & \textbf{\% Sign Changes}\\
\midrule
Height & 6,698 & 12\% \\
BMI    & 103  & 0.3\% \\
CHD    & 2  & $\leq$ 0.01\% \\
Asthma & 2  & $\leq$ 0.01\% \\
\bottomrule
\end{tabular}
\caption{\em Number of coefficient sign changes between the univariate and multivariate models for each phenotype and the percentage of coefficients that switched. UniLasso, by construction, eliminates all such sign reversals; every non-zero coefficient retains the sign of its univariate effect.}
\label{tab: sign_changes}
\end{table}

\section{Use of External Scores} \label{sec: external scores}

Our second contribution addresses the integration of external data to enhance predictive accuracy. In biomedical research, privacy regulations often restrict the sharing of raw individual-level data, meaning that researchers may only provide summary statistics rather than full datasets. \citet{uniLasso} proposed a method for incorporating external univariate scores from an independent dataset, denoted 
$\tt E$, into the uniLasso pipeline.

A collaborator from Finland provided external summary statistics from Finn-Gen, the Finnish counterpart to the UK Biobank, which comprises approximately 500,000 participants. The external scores from FinnGen consist of the results of univariate SNP regressions that include age, sex, and the first ten principal components as covariates.

We begin by computing the univariate intercepts and slopes $(\hat\beta_{0j}, \hat\beta_j)$, as in Step 1 of uniLasso. However, rather than relying on the standard formulation that uses leave-one-out (LOO) coefficients, \citet{uniLasso} propose an alternative regression framework, presented as equation (3.3) in their paper:

\begin{equation}  \label{eq:external scores}
\argmin_{\gamma} \sum_{i=1}^n (y_i - \gamma_0
-\sum_{j=1}^px_{ij}\gamma_j)^2 + \lambda\sum_{j=1}^p\frac{ |\gamma_j|}{|\Tilde\beta_j|}\quad
\mbox{ s.t. $\sign(\gamma_j)=\sign(\Tilde{\beta_j})\;\forall j$}
\end{equation}

The idea is to use the univariate coefficients from $\tt E$ rather than compute the LOO estimates from the training data. In the equation above these correspond to the $\Tilde{\beta}_j$ terms. Specifically, the magnitudes $|\Tilde\beta_j|$ serve as adaptive weights in the $\ell_1$-penalty term, while the signs of $\Tilde{\beta}_j$ agree with the signs of the estimated coefficients $\gamma_j$ (or they are zero). We draw the univariate slopes and intercepts from the UK Biobank training data, but use the external coefficients $\tilde{\beta}_j$ to define the adaptive penalty weights and sign constraints in the second-stage regression. The final model can be written as $\hat{\eta}(x) = \hat\gamma_0 + \sum_{j=1}^p \hat\gamma_j x_j.$
Note this alternative construction holds for binary responses as well.

The results for uniLasso with external scores (denoted uniLasso ES) are displayed alongside those of Lasso and standard uniLasso in Figure \ref{fig:test errors}. uniLasso ES achieves the highest predictive performance across all phenotypes evaluated except for a negligible 0.001 difference with Lasso for BMI. Compared to uniLasso, it appears that external scores will always boost the predictive power of the model. Although the resulting models are slightly less sparse than those from standard uniLasso, they remain consistently sparser than Lasso (again, see Table \ref{tab:nonzero_coefs}). The solution set for CHD is also very nearly 50\% sparser than Lasso in this framework. These findings demonstrate that incorporating external scores enhances both accuracy and interpretability, yielding models that outperform the ubiquitous Lasso in predictive power and sparsity.


\begin{table}[t]
\centering
\begin{tabular}{ccccc}
\toprule
\textbf{Similarity} &
\textbf{Height} &
\textbf{BMI} &
\textbf{CHD} &
\textbf{Asthma} \\
\midrule
$L \text{ and } U$     & 0.27 & 0.29  & 0.29  &  0.36 \\
$L \text{ and } E$     & 0.20 & 0.18  & 0.07  &  0.10 \\
$U \text{ and } E$     & 0.23 & 0.22  & 0.09  &  0.11 \\
\bottomrule
\end{tabular}
\caption{\em Proportion of shared nonzero coefficients between each pair of models (Lasso (L), uniLasso (U), and uniLasso with external scores (E)) for all phenotypes. Each value represents the size of the intersection divided by the size of the union of non-zero SNPs. Recall that the binary phenotypes have much smaller active sets, as shown in Table \ref{tab:nonzero_coefs}.}
\label{tab: similarity}
\end{table}

Table \ref{tab: sign_changes} shows that the logistic models exhibit very few sign changes. This naturally leads to examining how much alignment there is among the sets of selected SNPs across methods. Table \ref{tab: similarity} reports the sizes of each pairwise intersections divided by the size of their union. We examine the overlaps between Lasso, uniLasso, and uniLasso ES. Across all phenotypes, Lasso and uniLasso share the largest overlap, followed by Lasso and uniLasso ES. Lasso and uniLasso exhibit their largest similarity for the binary traits.

\section{Comparison of uniLasso to Other Approaches}
\label{sec: comparison summary}

In this section we summarize the performance differences among the methods considered. In addition to the three Lasso-based methods we analyze PRS-CS \citep{PRSCS}, a summary statistics-based Bayesian regression method. PRS-CS computes polygenic risk scores using GWAS summary statistics and an external LD reference panel, but it neither incorporates \textit{external} summary statistics nor induces sparsity in its effect-size estimates. For PRS-CS, we first generated GWAS summary statistics on the combined training and validation sets $(n = 269{,}711)$ with age, sex, and the top 10 PCs as covariates using PLINK v2.0a7 using MAF $\geq$ 0.0005 \citep{Chang2015PLINK2, PLINK2}. We then applied PRS-CS with default settings using the precomputed European-ancestry LD reference panel derived from the 1000 genomes samples \href{https://github.com/getian107/PRScs}{(PRScs GitHub Repo)}. Finally, we extracted the posterior effect-size estimates and computed polygenic risk scores using PLINK2’s \texttt{--score} subcommand \citep{Chang2015PLINK2}.

PRS-CS underperforms the three Lasso-based methods on all four of the phenotypes, though it nearly matches uniLasso ES on CHD (see Table \ref{tab:test_performance}). Combined with its higher computational cost and lack of sparsity, this limits its suitability for settings where interpretability is important. In particular, the dense set of nonzero effects makes it harder to gain intuition behind which variants are actually driving the signal. 

\begin{table}[t]
\centering
\begin{tabular}{lcccc}
\toprule
\textbf{Phenotype} & 
\textbf{Lasso} & 
\textbf{uniLasso} & 
\textbf{uniLasso ES} &
\textbf{PRS-CS} \\
\midrule
Height                  & 0.713 & 0.707 & \textbf{0.716} & 0.649 \\
BMI                     & \textbf{0.119} & 0.103 & 0.118 & 0.071 \\
Asthma                  & 0.619 & 0.620 & \textbf{0.623} & 0.597 \\
CHD                     & 0.758 & 0.757 & \textbf{0.760} & 0.759 \\
\bottomrule
\end{tabular}
\caption{\em Test set predictive performance of the four PRS methods for each phenotype. For Height and BMI, the metric is $R^2$; for Asthma and Coronary Heart Disease (CHD), the metric is AUC. uniLasso with external scores is denoted as uniLasso ES and PRS-CS is a competitive Bayesian regression method discussed in Section \ref{sec: comparison summary}.}
\label{tab:test_performance}
\end{table}

See Appendix \ref{app sec: runtimes} for a discussion regarding the difference in runtimes for each algorithm.

\section{Discussion}
\label{sec: discussion}
In this paper, we propose a new framework for analyzing ultra–high-dimensional omics data. Our approach builds on the univariate-guided sparse regression (uniLasso) algorithm recently introduced by \citet{uniLasso}. More broadly, our work demonstrates how uniLasso can be adapted into a practical framework for biobank-scale PRS construction and how external summary statistics can be integrated into such pipelines. Using the Python package \texttt{Adelie}, we implement polygenic risk score (PRS) algorithms for both continuous and binary outcomes. 

We evaluate our methods on SNP data to demonstrate their gains in sparsity and interpretability over existing approaches. Our experiments indicate that uniLasso achieves about 40\% more sparsity than Lasso while maintaining comparable predictive performance. Furthermore, integrating external data into the framework yields a noteworthy improvement in prediction accuracy. In the UK Biobank application, uniLasso with external scores (uniLasso ES) achieves the best predictive performance across all four measured phenotypes. Additionally, uniLasso ES retains substantially fewer non-zero coefficients than Lasso and far fewer than PRS-CS, which keeps all SNPs in the model. 

Across all settings, uniLasso’s enforced sparsity enables efficient computation in ultra–high-dimensional regimes while preserving, and often improving upon, the predictive accuracy of dense alternatives. It is particularly striking that for height, a canonical highly polygenic trait, our best uniLasso models rely on only about $34{,}000$ non-zero SNPs out of more than one million candidates, yet still achieve test $R^2 \approx 0.71$. Despite the extensive LD structure across the genome, uniLasso appears able to concentrate predictive signal onto a comparatively small subset of variants or LD blocks. This pattern is consistent with a view in which much of the SNP-heritability is concentrated in a modest number of genomic locations, even for traits that are often described as “highly polygenic”.

From this perspective, our results provide an empirical counterpoint to strictly “pangenomic” or infinitesimal models in which every variant is assumed to have a non-zero effect. Dense approaches like LDpred-inf and PRS-CS remain valuable, especially when only GWAS summary statistics are available, but our findings suggest that there is little predictive benefit in keeping millions of non-zero coefficients once LD and sample size are properly accounted for. A sparse-polygenic view, where a few tens of thousands of variants capture most of the predictive signal, appears sufficient for practical prediction in this setting.
Methods such as PRS-CS or infinitesimal-prior approaches effectively assume that every variant has a non-zero effect, but that most effects are very small. In the presence of LD, these dense models spread each signal over many correlated SNPs, so that each individual predictor carries a heavily shrunk, “flattened” effect size. This is attractive for aggregate prediction and some forms of inference, but it makes it difficult to identify which specific variants or LD blocks are truly driving the score. Moreover, for individual-level risk prediction, relying solely on correlated tag SNPs has a causal cost: whenever the LD between tag and causal variant is imperfect, individuals whose genotypes deviate from the dominant LD pattern receive contributions from non-causal predictors that are essentially noise. In contrast, a sparse multivariate method that concentrates weight on (or near) truly causal variants aims to reduce this mismatch between the mechanistic signal and the predictors used for stratification, potentially improving both interpretability and robustness of individual risk estimates.

More broadly, our comparisons highlight a generic tension in LD-structured high-dimensional regression. Traditional GWAS is purely univariate, so each SNP’s marginal effect conflates its own contribution with that of its correlated neighbors. Dense multivariate models that do not explicitly promote sparsity tend to share this signal across entire correlated sets, yielding many predictors with tiny, flattened effect sizes. In contrast, the $\ell_1$ penalty and sign constraints in uniLasso force the model to choose a subset of SNPs to carry the signal within each LD block, leading to larger, more interpretable per-variant effects without sacrificing out-of-sample prediction.

There are a number of associated challenges that we plan to pursue as a natural consequence of this work. The existing uniLasso framework handles $K = 2$-class targets but not $K > 2$. This is an inherently difficult problem since the $K$ coefficients for each feature are shift-invariant and hence their signs are not meaningful. We intend to develop and test new forms of uniLasso in the multiclass (multinomial) response context.
A second extension concerns applying uniLasso to sets of SNPs found from linkage disequilibrium (LD)-based pre-clustering (e.g., via clumping or pruning) and studying the resulting false discovery rate. There is recent work along these lines using Lasso which suggests this is a promising domain. Moreover, leave-one-chromosome-out (LOCO) cross-validation may help avoid signal leakage through LD and polygenic signals. Implementing LOCO folds within uniLasso could provide more reliable model assessment in genetic applications.

Finally, we are interested in developing a version of uniLasso that simultaneously incorporates both LOO fitted values and external univariate scores. As of now, uniLasso and uniLasso ES each leverage only one of these information sources; a unified approach that blends them could yield further gains in stability, sparsity, and predictive accuracy. Such a method would be especially valuable in large-scale genomic settings like the UK Biobank.

\bigskip

\subsubsection*{Acknowledgments}
We thank Etai Koronyo and Yosuke Tanigawa for their help and encouragement throughout this research process. T.H. was supported by NIH grant
R01 GM134483; R.T. was supported by NIH grant R01 GM134483 and NSF grant 19DMS1208164. M.A.R. is in part supported by National Human Genome Research Institute (NHGRI) under award R01HG010140, and by the National Institutes of Mental Health (NIMH) under award R01MH124244 both of the National Institutes of Health (NIH).

We will soon make available a Python package ``Lasso-PRS'' implementing both Lasso and uniLasso for polygenic risk scores.

\newpage

\appendix
\section*{Appendix}
\section{Algorithmic Runtimes} \label{app sec: runtimes}

\begin{table}[h]
\centering
\begin{tabular}{lcc}
\toprule
\textbf{Method / Stage} & \textbf{Cont. Response} & \textbf{Bin. Response} \\
\midrule
\multicolumn{3}{l}{\textit{Preprocessing}} \\
\quad uniLasso (pretraining) & 90 min & 4 hr \\
\quad PRS-CS (GWAS preprocessing) & 1 hr & 5 hr \\
\midrule
\multicolumn{3}{l}{\textit{Training Time}} \\
\quad Lasso & 2.5 hr & 85 min \\
\quad uniLasso (after pretraining) & 3 hr & 5 hr \\
\quad uniLasso ES & \textbf{1.5 hr} & \textbf{35 min} \\
\quad PRS-CS & 5.5 hr & 2.5 hr \\
\bottomrule
\end{tabular}
\caption{\em Approximate computation times for model training across all methods for continuous and binary phenotypes. Standard uniLasso includes an additional preprocessing stage in which the the LOO fitted values and univariate coefficients are computed for each feature. PRS-CS also requires a GWAS as input, which we count as preprocessing.}
\label{tab: timings}
\end{table}

An important practical consideration is the training time required for each algorithm, summarized in Table \ref{tab: timings}. All computations were performed on Stanford University’s high-performance computing cluster, Sherlock. For continuous outcomes, the preprocessing stage of uniLasso, which computes the leave-one-out (LOO) fitted values and the univariate coefficients, takes roughly 90 minutes on 64 CPUs for a dataset of size roughly about $250{,}000$ individuals and $710{,}000$ SNPs. For binary responses, the preprocessing runtime increases to around 4 hours due to the additional iterative approximation required in the IRLS optimization.

Fitting the uniLasso model with Gaussian loss after the computation of the pretraining values required approximately 3.5 hours. This is slightly longer than the average of 2.5 hours for Lasso, which in turn is greater than the 1.5 hour average for uniLasso with external scores (uniLasso ES). The reduced runtime of uniLasso ES arises from its formulation, which preserves the sparsity of the SNP matrix rather than populating it with dense LOO fitted values, as in the standard uniLasso implementation.

Training becomes even quicker when modeling binary outcomes. Under the same parameter configurations, Lasso finishes fitting in approximately 85 minutes, while uniLasso ES converges in only about 35 minutes. This exceptionally fast runtime further highlights the advantages of the uniLasso ES formulation. On the other hand, standard uniLasso requires an additional 5.5 hours of training after computing the necessary univariate and LOO estimates. With appropriate computational resources, for example GPU acceleration, this time could be substantially reduced. This is a direction we are actively pursuing. Nonetheless, the longer runtime is offset by the interpretability and unprecedented sparsity of the resulting model.

Across all response types, uniLasso with external scores achieves the fastest average training time among sparsity inducing competitors. This highlights the advantage of leveraging external information within the uniLasso framework, which reduces the need for intensive pretraining computations.

The PRS-CS pipeline involves substantial preprocessing and data formatting before the main model can be run, and additional steps are required afterward to aggregate the scores into a usable test metric. On Stanford’s Sherlock cluster, the primary PRS-CS Python command required roughly 5.5 hours for continuous phenotypes and 2.5 hours for binary phenotypes (Table \ref{tab: timings}). PRS-CS requires GWAS summary statistics as input, however, so we first run a genome-wide association study to generate the necessary single-variant effect estimates. This preprocessing stage showed the opposite pattern: about 1 hour for continuous responses versus 5 hours for binary. Overall, the end-to-end runtimes were similar, totaling approximately 6.5 hours and 7.5 hours for continuous and binary phenotypes respectively. In our experiments it showed somewhat weaker predictive performance compared to the Lasso-based methods, and does not naturally yield sparse models.

\newpage
\bibliography{references}

\begin{thebibliography}{12}
\providecommand{\natexlab}[1]{#1}
\providecommand{\url}[1]{\texttt{#1}}
\expandafter\ifx\csname urlstyle\endcsname\relax
  \providecommand{\doi}[1]{doi: #1}\else
  \providecommand{\doi}{doi: \begingroup \urlstyle{rm}\Url}\fi

\bibitem[Bycroft et~al.(2018)Bycroft, Freeman, Petkova, Band, Elliott, Sharp,
  Motyer, Vukcevic, Delaneau, O'Connell, Cortes, Welsh, Young, Effingham,
  McVean, Leslie, Allen, Donnelly, and Marchini]{UKBiobank}
Clare Bycroft, Colin Freeman, Dasha Petkova, Gavin Band, Lloyd~T. Elliott,
  Kevin Sharp, Allan Motyer, Damjan Vukcevic, Olivier Delaneau, Jared
  O'Connell, Adrian Cortes, Simon Welsh, Aaron Young, Michelle Effingham, Gil
  McVean, Stephen Leslie, Naomi Allen, Peter Donnelly, and Jonathan Marchini.
\newblock The uk biobank resource with deep phenotyping and genomic data.
\newblock \emph{Nature}, 562:\penalty0 203--209, 2018.
\newblock \doi{10.1038/s41586-018-0579-z}.
\newblock URL \url{https://doi.org/10.1038/s41586-018-0579-z}.

\bibitem[Chang et~al.(2015)Chang, Chow, Tellier, Vattikuti, Purcell, and
  Lee]{Chang2015PLINK2}
Christopher~C. Chang, Carson~C. Chow, Laurent C.~A. Tellier, Shashaank
  Vattikuti, Shaun~M. Purcell, and James~J. Lee.
\newblock Second-generation plink: rising to the challenge of larger and richer
  datasets.
\newblock \emph{GigaScience}, 4\penalty0 (1), 2015.
\newblock \doi{10.1186/s13742-015-0047-8}.

\bibitem[Chatterjee et~al.(2025)Chatterjee, Hastie, and Tibshirani]{uniLasso}
Sourav Chatterjee, Trevor Hastie, and Robert Tibshirani.
\newblock Univariate-guided sparse regression.
\newblock \emph{Harvard Data Science Review}, 7\penalty0 (3), 2025.
\newblock \doi{10.1162/99608f92.c79ff6db}.

\bibitem[DeBoever et~al.(2018)DeBoever, Tanigawa, Lindholm, McInnes, Lavertu,
  Ingelsson, Chang, and Ashley]{subset_UKBiobank}
Christopher DeBoever, Yosuke Tanigawa, Meaghan~E. Lindholm, Glen McInnes,
  Amanda Lavertu, Erik Ingelsson, Christopher Chang, and Euan~A. Ashley.
\newblock Medical relevance of protein-truncating variants across 337,205
  individuals in the {UK} biobank study.
\newblock \emph{Nature Communications}, 9:\penalty0 1612, 2018.
\newblock \doi{10.1038/s41467-018-03910-9}.
\newblock URL \url{https://doi.org/10.1038/s41467-018-03910-9}.

\bibitem[Friedman et~al.(2010)Friedman, Hastie, and Tibshirani]{glmnet}
Jerome Friedman, Trevor Hastie, and Robert Tibshirani.
\newblock Regularization paths for generalized linear models via coordinate
  descent.
\newblock \emph{Journal of Statistical Software}, 33\penalty0 (1):\penalty0
  1--22, 2010.
\newblock URL \url{https://www.jstatsoft.org/article/view/v033i01}.

\bibitem[Ge et~al.(2019)Ge, Chen, Ni, Feng, and Smoller]{PRSCS}
Tian Ge, Chia{-}Yi Chen, Ying Ni, Yung{‐}Chen~Austin Feng, and Jordan~W.
  Smoller.
\newblock Polygenic prediction via bayesian regression and continuous shrinkage
  priors.
\newblock \emph{Nature Communications}, 10\penalty0 (1):\penalty0 1776, 2019.
\newblock \doi{10.1038/s41467-019-09718-5}.

\bibitem[Purcell and Chang(2020)]{PLINK2}
Shaun Purcell and Christopher Chang.
\newblock \emph{PLINK 2.0}, 2020.
\newblock Available at \url{https://www.cog-genomics.org/plink/2.0/}.

\bibitem[Qian et~al.(2020)Qian, Tanigawa, Du, Aguirre, Chang, Tibshirani,
  Rivas, and Hastie]{Qian2020}
Junyang Qian, Yosuke Tanigawa, Wenfei Du, Matthew Aguirre, Chris Chang, Robert
  Tibshirani, Manuel~A. Rivas, and Trevor Hastie.
\newblock A fast and scalable framework for large-scale and
  ultrahigh-dimensional sparse regression with application to the uk biobank.
\newblock \emph{PLOS Genetics}, 16\penalty0 (10):\penalty0 e1009141, 2020.
\newblock \doi{10.1371/journal.pgen.1009141}.

\bibitem[Rad and Maleki(2020)]{RadMaleki2020}
Kaveh~R. Rad and Arian Maleki.
\newblock A scalable estimate of the out-of-sample prediction error via
  approximate leave-one-out cross-validation.
\newblock \emph{Journal of the Royal Statistical Society: Series B (Statistical
  Methodology)}, 82\penalty0 (4):\penalty0 965--996, 2020.
\newblock \doi{10.1111/rssb.12374}.
\newblock URL \url{https://doi.org/10.1111/rssb.12374}.

\bibitem[Tibshirani(1996)]{Tibs1996}
Robert Tibshirani.
\newblock Regression shrinkage and selection via the lasso.
\newblock \emph{Journal of the Royal Statistical Society: Series B
  (Methodological)}, 58\penalty0 (1):\penalty0 267--288, 1996.

\bibitem[Tusher et~al.(2001)Tusher, Tibshirani, and Chu]{Tusher2001}
V.~G. Tusher, R.~Tibshirani, and G.~Chu.
\newblock Significance analysis of microarrays applied to the ionizing
  radiation response.
\newblock \emph{Proceedings of the National Academy of Sciences}, 98\penalty0
  (9):\penalty0 5116--5121, 2001.
\newblock \doi{10.1073/pnas.091062498}.
\newblock URL \url{https://doi.org/10.1073/pnas.091062498}.

\bibitem[Yang and Hastie(2024)]{adelie2024}
James Yang and Trevor Hastie.
\newblock Adelie: A fast and flexible python package for group lasso and
  elastic net, 2024.
\newblock URL \url{https://github.com/JamesYang007/adelie}.
\newblock Accessed: 2025-10-09.

\end{thebibliography}

\end{document}